\documentclass[pra,aps,amsmath,amssymb,eqsecnum,showkeys]{revtex4}

\newcommand{\hh}{{\mathcal{H}}}
\newcommand{\lnp}{{\mathcal{L}}}
\newcommand{\lsp}{{\mathcal{L}}_{+}}
\newcommand{\pen}{\openone}
\newcommand{\Tr}{{\mathrm{Tr}}}
\newcommand{\bro}{{\boldsymbol{\rho}}}
\newcommand{\vbro}{{\boldsymbol{\varrho}}}
\newcommand{\bsg}{{\boldsymbol{\sigma}}}

\newcommand{\cle}{{\mathcal{E}}}
\newcommand{\clf}{{\mathcal{F}}}

\newcommand{\mc}{{\mathcal{M}}}

\newcommand{\clp}{{\mathcal{P}}}
\newcommand{\clq}{{\mathcal{Q}}}
\newcommand{\am}{{\mathsf{A}}}
\newcommand{\bn}{{\mathsf{B}}}

\newcommand{\mm}{{\mathsf{M}}}

\newcommand{\ppm}{{\mathsf{P}}}
\newcommand{\qpm}{{\mathsf{Q}}}
\newcommand{\az}{{\mathsf{Z}}}
\newcommand{\ax}{{\mathsf{X}}}
\newcommand{\lasf}{{\mathsf{\Lambda}}}
\newcommand{\iu}{{\mathtt{i}}}

\begin{document}
\clearpage
\preprint{}

\title{Separability conditions based on local fine-grained uncertainty relations}

\author{Alexey E. Rastegin}

\affiliation{Department of Theoretical Physics, Irkutsk State University,
Gagarin Bv. 20, Irkutsk 664003, Russia}

\begin{abstract}
Many protocols of quantum information processing use entangled
states. Hence, separability criteria are of great importance. We
propose new separability conditions for a bipartite
finite-dimensional system. They are derived by using fine-grained
uncertainty relations. Fine-grained uncertainty relations can be
obtained by consideration of the spectral norms of certain
positive matrices. One of possible approaches to separability
conditions is connected with upper bounds on the sum of maximal
probabilities. Separability conditions are often formulated for
measurements that have a special structure. For instance, mutually
unbiased bases and mutually unbiased measurements can be utilized
for such purposes. Using resolution of the identity for each
subsystem of a bipartite system, we construct some resolution of
the identity in the product of Hilbert spaces. Separability
conditions are then formulated in terms of maximal probabilities
for a collection of specific outcomes. The presented conditions
are compared with some previous formulations. Our results are
exemplified with entangled states of a two-qutrit system.
\end{abstract}


\maketitle

\pagenumbering{arabic}
\setcounter{page}{1}

\section{Introduction}\label{sec1}

The concept of entanglement plays a principal role in foundations
and applications of quantum mechanics. Entangled states provide a
tool for basic protocols of quantum information processing
\cite{nielsen}. The quantum parallelism of Deutsch
\cite{deutsch85} cannot be implemented without the use of
entangled states. Studies of quantum entanglement has a long
history. An existence of purely quantum correlations was first
emphasized in the Schr\"{o}dinger ``cat paradox'' paper
\cite{cat35} and in the Einstein--Podolsky--Rosen paper
\cite{epr35}. A role of such correlations is brightly shown in
specified experiments similar to Bohm's version of the EPR
experiment \cite{bohm51}. Historically, studies of these
conceptual questions not only provide better understanding
foundations of quantum physics \cite{az99}. Today, we see active
development of technological applications of rather sophisticated
quantum-mechanical effects.

Due to progress in quantum information processing, both the
detection and quantification of entanglement are very important.
Despite of many efforts, these problems are still the subject of
active research \cite{horodecki96,hhhh09}. For entanglement
detection, we may use conditions that are satisfied by all
separable states. The violation of such conditions will be
sufficient for detection. There exist several criteria of
entanglement. Among them, the positive partial transpose (PPT)
criterion \cite{peres96} and the reduction criterion
\cite{mhph1999} are well known. Detection of entanglement beyond
the PPT criterion can sometimes be realized with the realignment
criterion \cite{wu2003} or the computable cross-norm (CCN)
criterion \cite{rudolph05}. The criteria mentioned above are
formulated in terms of transformations of the given density
matrix.

Some separability conditions are immediately connected with
special measurement schemes. As was shown in a series of works
\cite{guhne04,giovan2004,devic05,guhne06,huang2010}, separability
conditions can be based on uncertainty relations of various forms.
The author of \cite{devic07} pointed out connections of such
criteria with the correlation matrix criterion. Recently,
entanglement properties were studied in the context of classical
correlations between outcomes of complementary measurements
\cite{mdm2015}. Since the Heisenberg famous paper \cite{wh27}
appeared, much many studies of uncertainty relations were
accomplished \cite{lahti}. Traditionally, uncertainty relations
are formulated in terms of the standard deviations within the
Robertson approach \cite{robert}. Entropic uncertainty relations
are currently the subject of active research \cite{ww10,brud11}.
Entropic formulations of the uncertainty principle allowed to
overcome some doubts connected with the traditional approach
\cite{deutsch,maass}.

Uncertainty relations of the Landau--Pollak type differ from
entropic ones. Although the original formulation of Landau and
Pollak \cite{lanpol61} is related to signal analysis, it can be
treated quantum mechanically \cite{maass}. Entropic-like
uncertainty relations based on this approach were considered in
\cite{zozor1311}. The Landau--Pollak uncertainty relation can be
treated as an example of fine-grained uncertainty relations. Such
relations have been proposed and motivated in \cite{oppwn10}.
Indeed, entropic bounds cannot distinguish the uncertainty
inherent in obtaining a particular combination of the outcomes
\cite{oppwn10}. Fine-grained uncertainty relations for some
special quantum measurements were derived in
\cite{renf13,rastqip15}. Using the Landau--Pollak uncertainty
relation, the authors of \cite{devic05} examined separability
conditions for a bipartite system of qubits. So, it is natural to
ask for separability conditions based on fine-grained uncertainty
relations.

Separability conditions can be derived for a scheme with local
measurements of the special type. Using measurements with a
complete set of mutually unbiased bases, the authors of
\cite{shbah12} proposed the so-called correlation measure for
entanglement detection. Similarly, the entanglement detection can
be realized with a symmetric informationally complete measurement
\cite{rast13b}. The correlation measure can also be introduced
with mutually unbiased measurements \cite{fei14,rastosid} and with
a general symmetric informationally complete measurement
\cite{feili14}. The latter is based on the exact purity-based
expression for the sum of squared probabilities \cite{rastpsic}.
Mutually unbiased measurements have been proposed as a weaker
version of mutually unbiased bases \cite{kagour}. Similarly,
general symmetric informationally complete measurements are an
extension of usual symmetric informationally complete measurements
\cite{kgour13}.

In the present paper, we will obtain separability conditions based
on local fine-grained uncertainty relations. Such conditions are
formulated in terms of maximal probabilities for a collection of
measurements. To build a resolution of the identity in the
product of Hilbert spaces, we use mutually unbiased bases or
mutually unbiased measurements on each of subsystems of a
bipartite system. The paper is organized as follows. In Section
\ref{sec2}, we introduce the used notation and recall some
required facts. In Section \ref{sec3}, separability conditions are
derived from fine-grained uncertainty relations. Some properties
of maximal probabilities in a measurement on the total system are
previously discussed. In Section \ref{sec4}, the presented
separability conditions are exemplified with some types of
entangled states of a bipartite system of qutrits. In Section
\ref{sec5}, we conclude the paper with a summary of results.

\section{Preliminaries}\label{sec2}

In this section, we recall some definitions and fix the notation.
Let $\lnp(\hh)$ be the space of linear operators on
$d$-dimensional Hilbert space $\hh$. By $\lsp(\hh)$, we denote the
set of positive semi-definite operators on $\hh$. A state of the
quantum system is described by the density matrix
$\bro\in\lsp(\hh)$ such that $\Tr(\bro)=1$. Let us recall also the
notion of the spectral norm. To each $\ax\in\lnp(\hh)$, we assign
$|\ax|\in\lsp(\hh)$ as the positive square root of
$\ax^{\dagger}\ax$. Eigenvalues of $|\ax|$ counted with their
multiplicities are the singular values of $\ax$ denoted as
$s_{i}(\ax)$. The spectral norm of $\ax$ is defined as
\begin{equation}
\|\ax\|_{\infty}:=\max\bigl\{s_{i}(\ax):{\,}1\leq{i}\leq{d}\bigr\}
{\>} . \label{spndf}
\end{equation}
Other norms are widely used for obtaining of quantitative measures in
quantum information theory \cite{watrous1}. Many of them are
defined in terms of singular values. Relations for such norms with
applications to quantum entropies of a composite system were
examined in \cite{rcun12}.

In quantum theory, measurements play a key role in the sense that
without them the mathematical formalism would have no physical
meaning. A measurement stage is one of central questions in
quantum information processing. Quantum measurements are generally
described within the POVM formalism \cite{peresq}. Let
$\mc=\{\mm_{i}\}$ be a set of elements $\mm_{i}\in\lsp(\hh)$ that
satisfy the completeness relation
\begin{equation}
\sum_{i\in\Omega(\mc)}{\mm_{i}}=\pen
\ . \label{cmprl}
\end{equation}
Here, $\pen$ is the identity operator on $\hh$, and $\Omega(\mc)$
is the set of possible measurement outcomes. For the
pre-measurement state $\bro$, the probability of $i$-th outcome is
represented as \cite{peresq}
\begin{equation}
p_{i}(\mc|\,\bro)=\Tr(\mm_{i}\bro)
\ . \label{promi}
\end{equation}
For the given $\mc$ and $\bro$, the maximal probability will be
denoted by $p_{\max}(\mc|\bro)$. To estimate a state of the given
quantum system, the observer can adopt several complementary
experiments. For instance, the spin-$1/2$ system can be dealt with
measurements of the three orthogonal components of spin
\cite{brz99}. Eigenbases of the three Pauli matrices give an
example of a complete set of mutually unbiased bases.

Let $\cle=\bigl\{|e_{i}\rangle\bigr\}$ and
$\clf=\bigl\{|f_{j}\rangle\bigr\}$ be two orthonormal bases in a
$d$-dimensional Hilbert space $\hh$. They are said to be mutually
unbiased if and only if for all $i$ and $j$,
\begin{equation}
\bigl|\langle{e}_{i}|f_{j}\rangle\bigr|=\frac{1}{\sqrt{d}}
\ . \label{twbs}
\end{equation}
Several orthonormal bases form a set of mutually unbiased bases
(MUBs), when each two of them are mutually unbiased. Mutually
unbiased bases have found use in many questions of quantum
information theory (see \cite{bz10} and references therein). The
states within MUBs are indistinguishable in the following sense.
If the two observables have unbiased eigenbases, then the
measurement of one observable reveals no information about
possible outcomes of the measurement of other. This property is
utilized in some schemes of quantum key distribution. When $d$ is
a prime power, we certainly have a construction of $d+1$ MUBs
\cite{bz10}. It is based on properties of prime powers and
corresponding finite fields \cite{wf89,kr04}. In general, however,
the maximal number of MUBs in $d$ dimensions is still an open
question \cite{bz10}. Approaches to MUBs beyond prime
power dimensionalities are considered in \cite{bbrv02,wb05}.

It is possible to formulate unbiased measurements under weaker
requirements. The authors of \cite{kagour} introduced mutually
unbiased measurements (MUMs). They show that a complete set of
$d+1$ measurements exists for all $d$. Let us consider two POVM
measurements $\clp=\{\ppm_{i}\}$ and $\clq=\{\qpm_{j}\}$. Each of
them contains $d$ elements such that
\begin{align}
& \Tr(\ppm_{i})=\Tr(\qpm_{j})=1
\ , \label{tmn1}\\
& \Tr(\ppm_{i}\qpm_{j})=\frac{1}{d}
\ . \label{dmn1}
\end{align}
The POVM elements are all of trace one, but now not of rank one.
The formula (\ref{dmn1}) replaces the square of (\ref{twbs}).
Further, two different elements from the same POVM obey
\begin{equation}
\Tr(\ppm_{i}\ppm_{j})=\delta_{ij}{\,}\varkappa
+(1-\delta_{ij}){\>}\frac{1-\varkappa}{d-1}
\ , \label{mjmk}
\end{equation}
where $\varkappa$ is the efficiency parameter \cite{kagour}. The
same condition is imposed on the elements of $\clq$. The
efficiency parameter characterizes a closeness of the POVM
elements to rank-one projectors \cite{kagour}. This parameter
satisfies \cite{kagour}
\begin{equation}
\frac{1}{d}<\varkappa\leq1
\ . \label{vklm}
\end{equation}
The case $\varkappa=1/d$ should be excluded, as it gives
$\ppm_{i}=\pen/d$ for all $i$. The value $\varkappa=1$, if
possible, leads to the standard case of mutually unbiased bases.
We can only say that the maximal efficiency is reached for prime
power $d$. More precise bounds on $\varkappa$ depend on an
explicit construction of POVM elements \cite{kagour}. The
Brukner--Zeilinger approach to quantum information can be realized
with MUMs instead of MUBs \cite{rastproca}.

We will study separability conditions that can be derived from
fine-grained uncertainty relations. In this regard, one result of
the paper \cite{imai07} should be recalled. In our notation, it is
written as follows. Let $\bigl\{\mc^{(1)},\ldots,\mc^{(N)}\bigr\}$
be a set of $N$ POVMs, and let some index
$i(t)\in\Omega(\mc^{(t)})$ be assigned to each $t=1,\ldots,N$. For
arbitrary $\bro$, we have \cite{imai07}
\begin{equation}
\sum_{t=1}^{N}{p_{i(t)}(\mc^{(t)}|\,\bro)}
\leq
1+\biggl(\sum_{s\neq{t}}{\Bigl\|\sqrt{\mm_{i(s)}^{(s)}}\sqrt{\mm_{i(t)}^{(t)}}\Bigr\|_{\infty}^{2}}\biggr)^{\!1/2}
. \label{mim6}
\end{equation}
This upper bound generalizes a weak version of the Landau--Pollak
uncertainty relation to the case of more than two measurements.
Recall that Landau and Pollak prove their uncertainty relation in
the context of signal analysis. Reformulation in
quantum-mechanical terms was mentioned in \cite{maass}. The
authors of \cite{devic05} discussed separability conditions based
on a weak version of the Landau--Pollak relation. In general, the
use of (\ref{mim6}) is not very immediate since the sum in its
right-hand side demands additional calculation or estimation. For
a set of MUBs, however, this sum is exactly calculated.

We now consider a bipartite system of two $d$-dimensional
subsystems. Its Hilbert space is the product
$\hh_{AB}=\hh_{A}\otimes\hh_{B}$ of two spaces
$\hh_{A}$ and $\hh_{B}$. Let us choose the orthonormal basis
$\bigl\{|i_{S}\rangle\bigr\}$, where $S=A,B$, for each of the two
spaces $\hh_{A}$ and $\hh_{B}$. A maximally entangled pure state
is then expressed as
\begin{equation}
|\Phi_{AB}^{+}\rangle=\frac{1}{\sqrt{d}}{\,}\sum_{i=1}^{d} |i_{A}\rangle\otimes|i_{B}\rangle
\ . \label{mest}
\end{equation}
Entangled states are a basic resource in quantum information.
Hence, the problem of efficient detection of entanglement is of
great importance \cite{hhhh09}. Let us recall basic notions
related to separability. A product state is any state of the form
$\bro_{A}\otimes\bro_{B}$ \cite{bengtsson}. When both the matrices
$\bro_{A}$ and $\bro_{B}$ are rank-one, we have a pure product
state. A bipartite mixed state is called separable, when its
density matrix $\bro_{AB}$ can be represented as a convex
combination of product states \cite{zhsl98}. That is, there exist
a probability distribution $\{q(n)\}$ and two sets
$\{\bro_{A}(n)\}$ and $\{\bro_{B}(n)\}$ such that
\begin{equation}
\bro_{AB}=\sum\nolimits_{n} q(n){\>}\bro_{A}(n)\otimes\bro_{B}(n)
\ . \label{sepsdf}
\end{equation}
Note that each separable state can also be expressed as a convex
combination of only pure product states. This fact easily follows
from (\ref{sepsdf}) by substitution of the corresponding spectral
decompositions. When representations of the form (\ref{sepsdf}) are
not possible, the state is called entangled \cite{zhsl98}.

\section{Separability conditions}\label{sec3}

In this section, we derive separability conditions from local
fine-grained uncertainty relations. Criteria based on local
uncertainty relations can be motivated as follows. Such criteria
are sometimes able to detect entanglement of states that escape
detection by both the CCN and PPT criteria \cite{guhne06}. To
realize entanglement detection, we will use MUBs and MUMs.
Fine-grained uncertainty relations for them were derived in
\cite{rastqip15}. Separability conditions will be formulated in
terms of maximal probabilities for POVMs on a total system. For a
bipartite system of qubits, the authors of \cite{devic05} gave
separability conditions in terms of maximal probabilities for
observables. Their result is based on lemma 1 of the paper
\cite{guhne04}. Before presenting separability conditions, we
shall formulate a similar statement about maximal probabilities in
measurements. This statement holds under an additional condition,
which we impose on a total POVM built of local ones. Let us begin
with the corresponding definition. We give it for a bipartite
system, since an extension to the multipartite case is obvious.

\newtheorem{prp31}{Definition}
\begin{prp31}\label{pon31}
Let $\clp=\{\ppm_{i}\}$ with $i\in\Omega(\clp)$ be a POVM in
$\hh_{A}$, and let $\clq=\{\qpm_{j}\}$ with $j\in\Omega(\clq)$ be
a POVM in $\hh_{B}$. Let subsets $\varpi(k)$ form a partition of
the Cartesian product $\Omega(\clp)\times\Omega(\clq)$ with the
following property. For each subset, two different pairs
$(i_{1},j_{1})\in\varpi(k)$ and $(i_{2},j_{2})\in\varpi(k)$ do not
intersect, namely $i_{1}\neq{i}_{2}$ and $j_{1}\neq{j}_{2}$. We
say that a POVM $\mc=\{\mm_{k}\}$ in $\hh_{A}\otimes\hh_{B}$ with
$k\in\Omega(\mc)$ is built of $\clp$ and $\clq$ according to the
family $\{\varpi(k)\}$, in signs $\mc(\clp,\clq)$, when
\begin{equation}
\mm_{k}=\sum_{(i,j)\in\varpi(k)}{\ppm_{i}\otimes\qpm_{j}}
\ . \label{lmkc}
\end{equation}
\end{prp31}

Note that a particular local value $i\in\Omega(\clp)$
($j\in\Omega(\clq)$) cannot appear twice or more in the same
subset of ordered pairs. We will also use this definition with two
orthonormal bases in $\hh_{A}$ and $\hh_{B}$, respectively.
Then the right-hand side of (\ref{lmkc}) involves the
corresponding rank-one projectors. Using probabilities, we
will focus on POVMs rather than on observables. The
following statement takes place.

\newtheorem{prp32}[prp31]{Proposition}
\begin{prp32}\label{pon32}
Let a POVM $\mc(\clp,\clq)$ in $\hh_{A}\otimes\hh_{B}$ be built of
local POVMs $\clp$ and $\clq$ according to Definition \ref{pon31}.
For all product states, we then have
\begin{align}
p_{\max}\bigl(\mc(\clp,\clq)\big|\,\bro_{A}\otimes\bro_{B}\bigr)
&\leq{p}_{\max}(\clp|\,\bro_{A})
\ , \label{pmlm}\\
p_{\max}\bigl(\mc(\clp,\clq)\big|\,\bro_{A}\otimes\bro_{B}\bigr)
&\leq{p}_{\max}(\clq|\,\bro_{B})
\ . \label{pmln}
\end{align}
For each separable state (\ref{sepsdf}), the quantity
$p_{\max}\bigl(\mc(\clp,\clq)\big|\,\bro_{AB}\bigr)$ is no greater
than
\begin{equation}
\underset{\bro_{A}(n)}{\max}\,p_{\max}\bigl(\clp|\,\bro_{A}(n)\bigl)
\ , \qquad
\underset{\bro_{B}(n)}{\max}\,p_{\max}\bigl(\clq|\,\bro_{B}(n)\bigl)
\ . \label{addab}
\end{equation}
\end{prp32}

{\bf Proof.} Since
$p_{i}(\clp|\,\bro_{A})\leq{p}_{\max}(\clp|\,\bro_{A})$ for all
$i\in\Omega(\clp)$, we write
\begin{align}
p_{k}\bigl(\mc(\clp,\clq)\big|\,\bro_{A}\otimes\bro_{B}\bigr)&=
\sum_{(i,j)\in\varpi(k)}{p_{i}(\clp|\,\bro_{A})\,p_{j}(\clq|\,\bro_{B})}
\nonumber\\
&\leq{p}_{\max}(\clp|\,\bro_{A})\sum_{j\in\theta(k)}{p_{j}(\clq|\,\bro_{B})}
\ . \label{lnmp}
\end{align}
By $\theta(k)$, we mean here the set of all those
$j\in\Omega(\clq)$ that appear in ordered pairs of $\varpi(k)$. As
any $j$ never appears twice or more, the normalization condition
completes the proof of (\ref{pmlm}). By a parallel argument, we
get (\ref{pmln}). Finally, the claim (\ref{addab}) follows from
(\ref{sepsdf}) and the linearity of the trace. $\blacksquare$

The statement of Proposition \ref{pon32} could be compared with
lemma 1 of the paper \cite{guhne04}. For a pair of local
observables with non-zero eigenvalues, there exists some
majorization relation between probability distributions
\cite{guhne04}. We refrain from presenting the details here, since
the results (\ref{pmlm}) and (\ref{pmln}) are quite sufficient for
our purposes. The formulas (\ref{pmlm}) and (\ref{pmln}) can be
violated, when subsets $\varpi(k)$ contain intersecting pairs. We
exemplify the claim with a system of two qubits, both in the
completely mixed state. Measuring each of qubits in some basis,
say $\bigl\{|0\rangle,|1\rangle\bigr\}$, we then obtain the
uniform distribution with two outcomes. For any local measurement,
the maximum probability is equal to $1/2$. Let us build a total
POVM with respect to the two subsets
$\bigl\{(0,0),(1,0),(1,1)\bigr\}$ and $\bigl\{(0,1)\bigr\}$. As
the former contains intersecting pairs, this measurement on the
two-qubit system does not share Definition \ref{pon31}. For the
first subset, we have the probability $3/4>1/2$. Thus, the
condition of Definition \ref{pon31} should be kept in constructing
POVMs of local measurements. In principle, each of the formulas
(\ref{pmlm}) and (\ref{pmln}) taken separately holds under weaker
conditions. Say, the former can be derived, when different pairs
of the same subset are allowed to intersect in their first entry
but not in the second. In the following, such measurements are not
used.

For entangled states, the statement of Proposition \ref{pon32}
does not hold in general. Let us consider a system of two qubits
in the pure state
\begin{equation}
|\Phi^{+}\rangle=\frac{|00\rangle+|11\rangle}{\sqrt{2}}
\ . \label{mestq}
\end{equation}
We further take the observable $\bsg_{z}\otimes\bsg_{z}$, where
$\bsg_{z}$ is diagonal in the basis
$\bigl\{|0\rangle,|1\rangle\bigr\}$. In the notation of Definition
\ref{pon31}, the measurement is based on the subsets
$\bigl\{(0,0),(1,1)\bigr\}$ and $\bigl\{(0,1),(1,0)\bigr\}$. As
the state (\ref{mestq}) is an eigenstate of
$\bsg_{z}\otimes\bsg_{z}$, the maximal probability is equal to
$1$. Further, the state of each of two qubits is completely mixed.
Measuring $\bsg_{z}$ on a single qubit will then result in the
uniform distribution with the maximal probability $1/2$. The
latter is less than for the combined system.

Together with Proposition \ref{pon32}, we will also use another
property of the maximal probability. For any POVM $\mc$ and a
convex combination of density matrices $\bro$ and $\vbro$, we have
\begin{equation}
p_{\max}\bigl(\mc\big|\lambda\bro+(1-\lambda)\vbro\bigr)
\leq\lambda\,p_{\max}(\mc|\,\bro)+(1-\lambda)\,p_{\max}(\mc|\,\vbro)
\ , \label{convb}
\end{equation}
where $\lambda\in[0;1]$. This property immediately follows from
(\ref{promi}) and the linearity of the trace. Our first collection of
separability conditions is posed as follows.

\newtheorem{prp33}[prp31]{Proposition}
\begin{prp33}\label{pon33}
Let $\bigl\{\cle^{(1)},\ldots,\cle^{(N)}\bigr\}$ be a set of $N$
MUBs in the space $\hh_{A}$, and let
$\bigl\{\clf^{(1)},\ldots,\clf^{(N)}\bigr\}$ be a set of $N$ MUBs
in the space $\hh_{B}$. Let $N$ POVMs
$\mc^{(t)}(\cle^{(t)},\clf^{(t)})$ in $\hh_{A}\otimes\hh_{B}$ be
constructed from these MUBs according to Definition \ref{pon31}.
If the density matrix $\bro_{AB}$ is separable, then
\begin{equation}
\sum_{t=1}^{N}{p_{\max}\bigl(\mc^{(t)}(\cle^{(t)},\clf^{(t)})\big|\,\bro_{AB}\bigr)}
\leq
\frac{N}{d_{S}}
\left(
1+\frac{d_{S}-1}{\sqrt{N}}
\right)
, \label{fngef}
\end{equation}
where $S=A,B$.
\end{prp33}

{\bf Proof.} The following statement was proved in
\cite{rastqip15}. Let $\bigl\{\cle^{(1)},\ldots,\cle^{(N)}\bigr\}$
be a set of MUBs in $d_{A}$-dimensional Hilbert space $\hh_{A}$.
Let some index $i(t)\in\Omega(\cle^{(t)})$ be assigned to each
$t=1,\ldots,N$. We then have
\begin{equation}
\sum_{t=1}^{N}{p_{i(t)}(\cle^{(t)}|\,\bro_{A})}
\leq
\frac{N}{d_{A}}
\left(
1+\frac{d_{A}-1}{\sqrt{N}}
\right)
. \label{fgmub}
\end{equation}
It is important here that the inequality (\ref{fgmub}) holds for
arbitrary choice of indices $i(t)$. Hence, we can replace
$p_{i(t)}(\cle^{(t)}|\,\bro_{A})$ with
$p_{\max}(\cle^{(t)}|\,\bro_{A})$ for all $t=1,\ldots,N$. For all
density matrices of the form (\ref{sepsdf}), we write
\begin{align}
&\sum_{t=1}^{N}{p_{\max}\bigl(\mc^{(t)}(\cle^{(t)},\clf^{(t)})\big|\,\bro_{AB}\bigr)}
\nonumber\\
&\leq\sum_{n}{q(n)}
\sum_{t=1}^{N}{p_{\max}\bigl(\mc^{(t)}(\cle^{(t)},\clf^{(t)})\big|\,\bro_{A}(n)\otimes\bro_{B}(n)\bigr)}
\label{gef1}\\
&\leq\sum_{n}{q(n)}
\sum_{t=1}^{N}{p_{\max}\bigl(\cle^{(t)}\big|\,\bro_{A}(n)\bigr)}
\ . \label{gef2}
\end{align}
Here, the step (\ref{gef1}) follows from (\ref{convb}), the step
(\ref{gef2}) follows from (\ref{pmlm}). Combining (\ref{gef2})
with (\ref{fgmub}) and the normalization condition
$\sum_{n}{q(n)}=1$, we obtain (\ref{fngef}) for $S=A$. Similarly,
we complete the proof for $S=B$. $\blacksquare$

The statement of Proposition \ref{pon33} provides necessary
conditions of separability. In principle, this result can be used
for bipartite systems combined of two quantum systems of different
dimensionality. It leads to a family of schemes of entanglement
detection with the use of mutually unbiased bases. Another
approach follows from the uncertainty relation (\ref{mim6}). For a
set of MUBs, the sum in the right-hand side of (\ref{mim6}) is
easily calculated. Indeed, the square root of any projector is
projector again, so that each summand becomes equal to $1/d$.
Using (\ref{mim6}) instead of (\ref{fgmub}) at the step
(\ref{gef2}) finally gives
\begin{equation}
\sum_{t=1}^{N}{p_{\max}\bigl(\mc^{(t)}(\cle^{(t)},\clf^{(t)})\big|\,\bro_{AB}\bigr)}
\leq
1+\sqrt{\frac{N^{2}-N}{d_{S}}}
\ , \label{fnmim6}
\end{equation}
where $S=A,B$ and $\bro_{AB}$ is separable. In general, the
condition (\ref{fnmim6}) seems to be weaker than (\ref{fngef}). On
the other hand, the condition (\ref{fnmim6}) may sometimes lead to
a better restriction. It will be exemplified below. Thus, both the
results (\ref{fngef}) and (\ref{fnmim6}) are of interest. In each
concrete case, we should choose more restrictive condition.

The violation of any of the relations (\ref{fngef}) and
(\ref{fnmim6}) implies that the measured state is entangled. To
increase an efficiency of detection, we try to use as many
different MUBs as possible. Various constructions of POVMs
$\mc^{(t)}(\cle^{(t)},\clf^{(t)})$ in $\hh_{A}\otimes\hh_{B}$ may
be considered. A general scheme for a bipartite system of two
$d$-dimensional subsystems will be described in the next section.
In the case of two subsystems of the same dimensionality, we will
omit subscript $S=A,B$ in the conditions (\ref{fngef}) and
(\ref{fnmim6}).

The authors of \cite{devic05} obtained separability conditions for
a bipartite system of qubits on the base of the Landau--Pollak
uncertainty relation. Their conditions are formulated in terms of
a tensor product of the Pauli observables. Substituting $N=2$ and
$d=2$ into (\ref{fngef}), for each separable $\bro_{AB}$ we have
\begin{equation}
\sum_{t=1}^{2}{p_{\max}(\mc^{(t)}|\,\bro_{AB})}
\leq1+\frac{1}{\sqrt{2}}\approx1.707
\ . \label{dev1}
\end{equation}
The right-hand side of (\ref{dev1}) also follows from the
Landau--Pollak uncertainty relation. Thus, the result
(\ref{fngef}) includes one of the separability conditions of
\cite{devic05} as a particular case. Our approach differs in the
following respects. First, it holds for all those POVMs that can
be built of local MUBs with respect to Definition \ref{pon31}.
Second, it is not formulated in terms of observables. Note also
that the bound (\ref{fngef}) is not tightest \cite{devic05,pkm14}.

When $N=3$ and $d=2$, for separable $\bro_{AB}$ the formula
(\ref{fngef}) gives
\begin{equation}
\sum_{t=1}^{3}{p_{\max}(\mc^{(t)}|\,\bro_{AB})}
\leq\frac{3}{2}
\left(
1+\frac{1}{\sqrt{3}}
\right)\approx2.366
\ . \label{dev2}
\end{equation}
Of course, for particular choices
of concrete POVM elements the general bound (\ref{dev2}) may be
improved. Indeed, even if we obtain some tight bound on the sum of
maximal probabilities for local measurements, we still do not have
a tight bound for the sum of maximal probabilities for the total
system. Particularly, a direct maximization could be used in
simple cases. The authors of \cite{devic05} provided an example of
three observables on a qubit pair, when the sum of three
probabilities is bounded from above by $2$. The latter was
obtained by performing a direct maximization of the sum of
probabilities in product states \cite{devic05}. For complementary
qubit observables, a similar approach was used in studying
entropic uncertainty and certainty relations
\cite{rastqip12,rastctp}. This direction was originally initiated
in the papers \cite{ivan92,sanchez93}. On the other hand, a direct
optimization becomes hardly appropriate with growth of the
dimensionality. The bound (\ref{dev2}) is a particular case of the
result (\ref{fngef}), which has been proved for arbitrary finite
$d$. It is rather natural that bounds of the form (\ref{fngef})
are not tight.

For prime power $d$, we can build a set of $d+1$ MUBs for each
subsystem. This case is apparently of most practical interest,
since in practice we will rather deal with systems of qubits or
qutrits. The condition (\ref{fngef}) then gives
\begin{equation}
\sum_{t=1}^{d+1}{p_{\max}\bigl(\mc^{(t)}(\cle^{(t)},\clf^{(t)})\big|\,\bro_{AB}\bigr)}
\leq
\frac{d+1}{d}
\left(
1+\frac{d-1}{\sqrt{d+1}}
\right)
, \label{fngef1}
\end{equation}
where $\bro_{AB}$ is separable. In the same case, the
condition (\ref{fnmim6}) reads
\begin{equation}
\sum_{t=1}^{d+1}{p_{\max}\bigl(\mc^{(t)}(\cle^{(t)},\clf^{(t)})\big|\,\bro_{AB}\bigr)}
\leq
1+\sqrt{d+1}
\ . \label{fnmim61}
\end{equation}
In this especially important case, the condition (\ref{fngef}) is
better than (\ref{fnmim6}). Indeed, for all $d\geq2$ the
right-hand side of (\ref{fngef1}) is strictly less than the
right-hand side of (\ref{fnmim61}). For $d=2$, the former is
approximately $2.366$, whereas the latter is approximately
$2.732$. A relative difference is more than $15$ \%. This
difference decreases with growth of $d$. For sufficiently large
$d$, the conditions (\ref{fngef1}) and (\ref{fnmim61}) are almost
the same. For two MUBs in the qubit case, when $N=2$ and $d=2$, we
respectively obtain $1+1/\sqrt{2}\approx1.707$ from (\ref{fngef})
and $2$ from (\ref{fnmim6}). A relative difference is more than
$17$ \%. For $N=2$ and sufficiently small $d$, the right-hand side
of (\ref{fngef}) is also less than the right-hand side of
(\ref{fnmim6}). With growth of $d$ at $N=2$, however, we observe
some domain in which the condition (\ref{fnmim6}) is stronger.
Here, a relative difference is firstly small, say, about $2$ \%
for $d=10$. This fact exemplifies that the conditions
(\ref{fngef}) and (\ref{fnmim6}) are both of interest. At the same
time, the result (\ref{fngef}) is more relevant with respect to
the case of most practical interest.

Except for prime power $d$, the maximal number of MUBs in $d$
dimensions is still an open question \cite{bz10}. Today, the
answer is unknown even for $d=6$. In this sense, alternate
approaches are of interest. The authors of \cite{kagour}
introduced the concept of mutually unbiased measurements. Elements
of such measurements are not of rank one. This approach with
lesser measurement efficiency is easy to construct. A complete set
of $d+1$ MUMs can be given explicitly for arbitrary finite $d$
\cite{kagour}. In the paper \cite{rastqip15}, we derived
fine-grained uncertainty relations for a set of MUMs. These
relations lead to a collection of separability conditions.

\newtheorem{prp34}[prp31]{Proposition}
\begin{prp34}\label{pon34}
Let $\bigl\{\clp^{(1)},\ldots,\clp^{(N)}\bigr\}$ be a set of $N$
MUMs of the efficiency $\varkappa_{A}$ in $\hh_{A}$, and let
$\bigl\{\clq^{(1)},\ldots,\clq^{(N)}\bigr\}$ be a set of $N$ MUMs
of the efficiency $\varkappa_{B}$ in $\hh_{B}$. Let $N$ POVMs
$\mc^{(t)}(\clp^{(t)},\clq^{(t)})$ in $\hh_{A}\otimes\hh_{B}$ be
constructed from these MUMs according to Definition
\ref{pon31}. If the density matrix $\bro_{AB}$ is separable, then
\begin{equation}
\sum_{t=1}^{N}{p_{\max}\bigl(\mc^{(t)}(\clp^{(t)},\clq^{(t)})\big|\,\bro_{AB}\bigr)}
\leq
\frac{N}{d_{S}}
\left(
1+\sqrt{\frac{(d_{S}-1)(\varkappa_{S}\,d_{S}-1)}{N}}
\right)
, \label{fngpq}
\end{equation}
where $S=A,B$.
\end{prp34}

{\bf Proof.} The following statement was proved in
\cite{rastqip15}. Let $\bigl\{\clp^{(1)},\ldots,\clp^{(N)}\bigr\}$
be a set of MUMs of the efficiency $\varkappa_{A}$ in
$d_{A}$-dimensional Hilbert space $\hh_{A}$. Let some index
$i(t)\in\Omega(\clp^{(t)})$ be assigned to each $t=1,\ldots,N$. We
then have
\begin{equation}
\sum_{t=1}^{N}{p_{i(t)}(\clp^{(t)}|\,\bro_{A})}
\leq
\frac{N}{d_{A}}
\left(
1+\sqrt{\frac{(d_{A}-1)(\varkappa_{A}\,d_{A}-1)}{N}}
\right)
. \label{fgmum}
\end{equation}
This inequality also holds for arbitrary choice of indices $i(t)$.
Remaining steps are written similarly to the proof of
Proposition \ref{pon33}. $\blacksquare$

The statement of Proposition \ref{pon34} generalizes (\ref{fngef})
to the case of mutually unbiased measurements. In the paper
\cite{rastqip15}, fine-grained uncertainty relations for MUMs were
based only on the formulas (\ref{tmn1}) and (\ref{dmn1}). Hence,
the separability condition (\ref{fngpq}) holds irrespectively to
the explicit construction for MUMs given in \cite{kagour}. Thus,
entanglement detection can be realized with mutually unbiased
measurements, at least in principle. We may use a complete set of
$d+1$ of MUMs for those $d$, for which $d+1$ MUBs are not
available. For the efficiency $\varkappa_{S}=1$, the condition
(\ref{fngpq}) is reduced to (\ref{fngef}). We may also ask for a
similar extension of (\ref{fnmim6}). In the case of MUMs, the sum
in the right-hand side of (\ref{mim6}) is difficult to evaluate.
At present, we can give only the condition (\ref{fngpq}) based on
the results of \cite{rastqip15}.

Finally, we shortly mention applications of the separability
conditions (\ref{fngef}), (\ref{fnmim6}), and (\ref{fngpq}) to
multipartite systems. Various approaches to study multipartite
entanglement are considered in
\cite{popescu98,toth06,huangy12,huber13,zhao15}. In general, the
problem becomes more complicated \cite{popescu98}. It can be
illustrated with the case of tripartite systems. Here, we should
distinguish between fully separable states and biseparable states
\cite{devic05}. Fully separable states are mixtures of product
states of the form $\bro_{A}\otimes\bro_{B}\otimes\bro_{C}$.
Biseparable states are mixtures of the form
$\bro_{A}\otimes\bro_{BC}$, in which $\bro_{BC}$ is not separable.
For multipartite system, Definition \ref{pon31} should be
reformulated accordingly. Then the conditions of this section can
be treated as biseparability conditions. For a multipartite
system, we will group the subsystems into two larger groups.
Further, the described schemes can be used. In principle, this
issue may be a theme of separate investigation.

\section{Applications to some states of a two-qutrit system}\label{sec4}

In this section, we apply some of the above separability
conditions to states whose separability limits are already known.
As separability of qubit systems are well studied in the
literature, we will focus on systems of qutrits. Such system are
also of great interest since the qutrit can be implemented by a
biphoton \cite{kwek02}. Bipartite separability conditions are
typically illustrated with density matrices of the form
\begin{equation}
(1-s)\vbro_{sep}+s\,|\Psi\rangle\langle\Psi|
\ . \label{werrd}
\end{equation}
Here, the density matrix $\vbro_{sep}$ is separable,
$|\Psi\rangle$ is a maximally entangled state, and $s\in[0;1]$.
Taking $\vbro_{sep}$ to be the completely mixed state, the form
(\ref{werrd}) leads to the bipartite case of Werner states
\cite{werner89}. That is, we consider mixtures of the
completely mixed state and a maximally entangled pure state. A
bipartite Werner state is separable if and only if
\cite{rubin2000}
\begin{equation}
s\leq\frac{1}{d+1}
\ . \label{crub}
\end{equation}
The authors of \cite{rubin2000} also gave a necessary and
sufficient condition for multipartite Werner states. For a
bipartite system of two qutrits, the inequality (\ref{crub}) reads
$s\leq1/4$. We will exemplify separability conditions of the
previous section by applying them to states (\ref{werrd}) of a
system of two qutrits.

We will use the result (\ref{fngef}) formulated for the scheme
with MUBs. Let us describe briefly our construction for arbitrary
$d$. In any base, the number index of its elements runs integers
from $0$ up to $d-1$. To each basis $\cle^{(t)}$, we assign the
operator $\am^{(t)}$. It is taken as diagonal in that basis and
represented as
\begin{equation}
\am^{(t)}={\mathrm{diag}}\bigl(1,\omega,\ldots,\omega^{d-1}\bigr)
\ , \label{amtdf}
\end{equation}
where $\omega=\exp(\iu2\pi/d)$ is a primitive root of the unit.
That is, each $\am^{(t)}$ is taken in own eigenbasis as the
corresponding generalized Pauli operator. Similarly, we assign the
operator $\bn^{(t)}$ to each basis $\clf^{(t)}$ so that it is
described by the right-hand side of (\ref{amtdf}) in this basis.
The spectrum of $\am^{(t)}\otimes\bn^{(t)}$ also contains $d$
powers of $\omega$. For all $t=1,\ldots,N$, we write
\begin{equation}
\am^{(t)}\otimes\bn^{(t)}=\sum_{k=0}^{d-1}{\omega^{k}\,\lasf_{k}^{(t)}}
\ . \label{ambnd}
\end{equation}
In terms of vectors of the bases $\cle^{(t)}$ and $\clf^{(t)}$,
one gives
\begin{equation}
\lasf_{k}^{(t)}=
\sum_{i=0}^{d-1}{|e_{i}^{(t)}\rangle\langle{e}_{i}^{(t)}|\otimes|f_{k\ominus{i}}^{(t)}\rangle\langle{f}_{k\ominus{i}}^{(t)}|}
\ , \label{lasdc}
\end{equation}
where the sign ``$\ominus$'' denotes the subtraction in
$\mathbb{Z}/d$. In the described scheme, we finally have
$\mc^{(t)}=\bigl\{\lasf_{k}^{(t)}\bigr\}$. For $d=2$, the
operators (\ref{amtdf}) and (\ref{ambnd}) are Hermitian. It is not
the case for $d>2$. In effect, the operators (\ref{amtdf}) and
(\ref{ambnd}) are only auxiliary in order to get the projection
operators (\ref{lasdc}). In a certain sense, the above scheme is
an extension of the case of two qubits considered in
\cite{devic05}.

In the case of qutrit system, we have $\omega=\exp(\iu2\pi/3)$.
Four MUBs in the Hilbert space of qutrit can be written as
\begin{align}
&\left\{
\begin{pmatrix}
1 \\
0 \\
0
\end{pmatrix},
{\>}
\begin{pmatrix}
0 \\
1 \\
0
\end{pmatrix},
{\>}
\begin{pmatrix}
0 \\
0 \\
1
\end{pmatrix}
\right\}
{\,},
&\left\{
\frac{1}{\sqrt{3}}
\begin{pmatrix}
1 \\
1 \\
1
\end{pmatrix},
{\>}
\frac{1}{\sqrt{3}}
\begin{pmatrix}
1 \\
\omega^{*} \\
\omega
\end{pmatrix},
{\>}
\frac{1}{\sqrt{3}}
\begin{pmatrix}
1 \\
\omega \\
\omega^{*}
\end{pmatrix}
\right\}
{\,}, \label{bas12}\\
&\left\{
\frac{1}{\sqrt{3}}
\begin{pmatrix}
1 \\
\omega \\
1
\end{pmatrix},
{\>}
\frac{1}{\sqrt{3}}
\begin{pmatrix}
1 \\
1 \\
\omega
\end{pmatrix},
{\>}
\frac{1}{\sqrt{3}}
\begin{pmatrix}
\omega \\
1 \\
1
\end{pmatrix}
\right\}
{\,},
&\left\{
\frac{1}{\sqrt{3}}
\begin{pmatrix}
1 \\
1 \\
\omega^{*}
\end{pmatrix},
{\>}
\frac{1}{\sqrt{3}}
\begin{pmatrix}
\omega^{*} \\
1 \\
1
\end{pmatrix},
{\>}
\frac{1}{\sqrt{3}}
\begin{pmatrix}
1 \\
\omega^{*} \\
1
\end{pmatrix}
\right\}
{\,}. \label{bas34}
\end{align}
Here, $\omega^{*}=\omega^{2}$ is the complex conjugation of
$\omega$. When one of MUBs is taken as the standard base, other
MUBs can be described in terms of complex Hadamard matrices. This
observation was used for classification of MUBs in low dimensions
\cite{bwb10}. The bases (\ref{bas12}) are respectively the
eigenbases of the generalized Pauli operators
\begin{equation}
\az=\begin{pmatrix}
1 & 0 & 0 \\
0 & \omega & 0 \\
0 & 0 & \omega^{*}
\end{pmatrix}
{\,},
\qquad
\ax=\begin{pmatrix}
0 & 0 & 1 \\
1 & 0 & 0 \\
0 & 1 & 0
\end{pmatrix}
{\,}. \label{azmatr}
\end{equation}
The bases (\ref{bas34}) are respectively the eigenbases of the
operators $\az\ax$ and $\az\ax^{2}$. In each base, the kets are
arranged according to the order of eigenvalues $1$, $\omega$,
$\omega^{*}$. It is easy to see that $\az\ax=\omega\ax\az$. The
bases (\ref{bas12})--(\ref{bas34}) can be used in studying
higher-dimensional quantum key distribution protocols
\cite{gjvwz06,mdg13}.

Detecting entanglement of states of the form (\ref{werrd})
generally depends on the choice of $|\Psi\rangle$ with respect to
the taken measurement projectors. For a system of two qutrits, we
consider four projective measurements corresponding to the normal
operators
\begin{equation}
\az\otimes\ax
\ , \qquad
\ax\otimes\az
\ , \qquad
\az\ax\otimes\az\ax
\ , \qquad
\az\ax^{2}\otimes\az\ax^{2}
\ . \label{forop}
\end{equation}
The first three operators are mutually commuting, as we see
from $\az\ax=\omega\ax\az$. Formulating separability conditions in
terms of maximal probabilities, these probabilities should be
sufficiently larger. For example, such conditions are useful, when
$|\Psi\rangle$ is a common eigenstates of the three commuting
operators. We further take a maximally entangled state as
\begin{equation}
|\Psi\rangle=\frac{1}{\sqrt{3}}
{\>}\bigl(
|z_{0}x_{0}\rangle+|z_{1}x_{2}\rangle+|z_{2}x_{1}\rangle
\bigr)
\, . \label{mqtr}
\end{equation}
By $|z_{i}\rangle$ and $|x_{j}\rangle$, we respectively mean the
eigenstates of $\az$ and $\ax$ so that the subscript shows the
power of $\omega$. Constructing the projection operators according
to the formulas (\ref{amtdf})--(\ref{lasdc}) gives the following.
To the operator $\az\otimes\ax$, we assign the three projectors
\begin{align}
\lasf_{0}^{(1)}&=|z_{0}x_{0}\rangle\langle{z}_{0}x_{0}|+
|z_{1}x_{2}\rangle\langle{z}_{1}x_{2}|+|z_{2}x_{1}\rangle\langle{z}_{2}x_{1}|
\ , \label{laf0}\\
\lasf_{1}^{(1)}&=|z_{0}x_{1}\rangle\langle{z}_{0}x_{1}|+
|z_{1}x_{0}\rangle\langle{z}_{1}x_{0}|+|z_{2}x_{2}\rangle\langle{z}_{2}x_{2}|
\ , \label{laf1}\\
\lasf_{2}^{(1)}&=|z_{0}x_{2}\rangle\langle{z}_{0}x_{2}|+
|z_{1}x_{1}\rangle\langle{z}_{1}x_{1}|+|z_{2}x_{0}\rangle\langle{z}_{2}x_{0}|
\ . \label{laf2}
\end{align}
They project a bipartite state on the eigenspaces, which
correspond to the eigenvalues $1$, $\omega$, $\omega^{*}$. For
other three bases, projectors on the product space are constructed
in a similar way. They can directly be obtained from the spectral
decomposition of the operators $\ax\otimes\az$,
$\az\ax\otimes\az\ax$, and $\az\ax^{2}\otimes\az\ax^{2}$. We will
denote the outcomes as the eigenvalues, though these operators are
not Hermitian. Recall that generalized Pauli operators are used
for convenience of describing the scheme
(\ref{amtdf})--(\ref{lasdc}).

Let us consider probability distributions for the pre-measurement
state $|\Psi\rangle$. In three of the all four cases, the state
(\ref{mqtr}) is an eigenstate, namely
\begin{equation}
(\az\otimes\ax)|\Psi\rangle=|\Psi\rangle
\ , \qquad
(\ax\otimes\az)|\Psi\rangle=|\Psi\rangle
\ , \qquad
(\az\ax\otimes\az\ax)|\Psi\rangle=\omega|\Psi\rangle
\ . \label{psien}
\end{equation}
In the three cases, we therefore obtain a deterministic
probability distribution. For the operator
$\az\ax^{2}\otimes\az\ax^{2}$, we have the uniform distribution
with three outcomes. This fact is easily derived
immediately.

Let us take $\vbro_{sep}$ to be the completely mixed state. In
each of the four cases, the three projectors has the same trace
equal to $3$. So, the completely mixed state always leads to the
uniform distribution for the three outcomes denoted by $1$,
$\omega$, $\omega^{*}$. Using these facts, we find the desired
maximal probabilities for the state (\ref{werrd}). The result
(\ref{fngef}) gives a collection of separability conditions. Since
the right-hand side of (\ref{fngef}) is equal to $1$ for $N=1$,
several measurements should be involved. To study entanglement of
the given state, we will rather use as more complementary
measurements as possible. In the case considered, we can use the
four MUBs (\ref{bas12})--(\ref{bas34}). Finding the maximal
probabilities for each measurement, we then check separability
conditions. For the state (\ref{werrd}) defined with the
completely mixed $\vbro_{sep}$ and (\ref{mqtr}), an efficiency of
entanglement detection is maximized with the three MUBs for which
$|\Psi\rangle$ leads to the maximal probability equal to $1$. For
$N=3$ and $d=3$, the right-hand side of (\ref{mqtr}) is equal to
$1+2/\sqrt{3}$. This upper bound can actually be improved. For any
three of the four bases (\ref{bas12})--(\ref{bas34}) and arbitrary
qutrit state $\bro$, one has the inequality \cite{rastqip15}
\begin{equation}
\sum_{t=1}^{3}{p_{\max}(\cle^{(t)}|\,\bro)}
\leq
1+\frac{2}{\sqrt{3}}{\>}
\cos\frac{\pi}{18}
\ . \label{fngef033}
\end{equation}
Using this bound instead of (\ref{fgmub}), we obtain
\begin{equation}
\sum_{t=1}^{3}{p_{\max}\bigl(\mc^{(t)}(\cle^{(t)},\clf^{(t)})\big|\,\bro_{AB}\bigr)}
\leq
1+\frac{2}{\sqrt{3}}{\>}
\cos\frac{\pi}{18}
\ . \label{fngef33}
\end{equation}
Hence, we have arrived at a conclusion. The separability
condition (\ref{fngef33}) detects entanglement when
\begin{equation}
s>\frac{1}{\sqrt{3}}{\>}
\cos\frac{\pi}{18}\approx0.569
\ . \label{sbon}
\end{equation}
Similarly, we can consider other forms of $\vbro_{sep}$, say, the
matrix
\begin{equation}
\frac{1}{3}
\,\Bigl(
|z_{0}z_{0}\rangle\langle{z}_{0}z_{0}|+|z_{1}z_{1}\rangle\langle{z}_{1}z_{1}|+|z_{2}z_{2}\rangle\langle{z}_{2}z_{2}|
\Bigr)
{\>}. \label{srsep}
\end{equation}
For all the four measurements, this density matrix also leads to
the uniform distribution with three outcomes. Thus, the
separability condition (\ref{fngef33}) again detects entanglement
when the parameter $s$ obeys (\ref{sbon}).

To reach more efficient criteria, we should perform a direct
maximization of the sum of probabilities in product states of a
pair of qutrit. This procedure is not obvious and rather
difficult. As we already mentioned, detection of entanglement of
the states (\ref{werrd}) within the considered scheme depends on
the choice of $|\Psi\rangle$. Let us consider the state
(\ref{werrd}), in which (\ref{mqtr}) is replaced with
\begin{equation}
|\Phi\rangle=\frac{1}{\sqrt{3}}
{\>}\bigl(
|z_{0}z_{0}\rangle+|z_{1}z_{1}\rangle+|z_{2}z_{2}\rangle
\bigr)
\ . \label{mqtr1}
\end{equation}
This ket is not an eigenstate of any of the operators
(\ref{forop}). In all the four measurements, the state
(\ref{mqtr1}) leads to the uniform distribution with three
outcomes. To reach entanglement detection with the use
of (\ref{fngef}), we should locally rotate the bases
(\ref{bas12})--(\ref{bas34}) with respect to the standard one.
This is a limitation of the scheme with
considering maximal probabilities. On the other hand, no
universal entanglement criteria are now known. Say, the PPT
criterion is necessary and sufficient for $2\times2$ and
$2\times3$ systems \cite{horodecki96}, but in higher dimensional
systems some entangled states escape detection. Thus, separability
conditions of the considered type can be useful, at least as
additional to other criteria.

\section{Conclusion}\label{sec5}

We have derived a collection of separability conditions for a
bipartite quantum system. The presented separability conditions
are obtained from local fine-grained uncertainty relations. They
are based on considering maximal probabilities for a set of
measurements. Separability criteria are often formulated for
measurements that have a special structure. For example, such
measurements can be built of mutually unbiased bases or mutually
unbiased measurements on subsystems. The considered schemes allow
a freedom in constructing a total measurement. Separable states
inevitably fulfill upper bounds that follow from local uncertainty
relations for a particular subsystem. Entangled states sometimes
violate such conditions. Separability conditions are obtained for
the measurement schemes based on MUBs as well as on MUMs. Of
course, we usually try to use as many different measurements as
possible. Actually, an efficiency of detection within the
described schemes depends on orientation of local measurement
bases and number of involved measurements. Main findings are
exemplified with some entangled states of a bipartite system of
two qutrits. Separability conditions of the considered type can be
used in entanglement detection together with other criteria.

\end{document}